\documentclass[journal]{IEEEtran} 
\IEEEoverridecommandlockouts
\usepackage{cite}
\usepackage{graphicx}
\usepackage{textcomp}
\usepackage[dvipsnames]{xcolor}
\usepackage{subcaption}
\usepackage{multirow}
\usepackage{colortbl}
\usepackage{adjustbox}
\usepackage{amsmath,amssymb,amsfonts,amsthm}
\usepackage[linesnumbered,ruled,vlined]{algorithm2e}
\usepackage[noend]{algpseudocode}
\usepackage{tikz}
\usepackage{dblfloatfix}
\usepackage{bigdelim} 
\usepackage{makecell}
\usepackage{arydshln}
\def\BibTeX{{\rm B\kern-.05em{\sc i\kern-.025em b}\kern-.08em
    T\kern-.1667em\lower.7ex\hbox{E}\kern-.125emX}}

\newcommand{\lite}{\texttt{LITE}}

\newcommand{\textbu}[1]{{\textbf{\underline{#1}}}}

\newcommand{\bstars}{\textsuperscript{$\bigstar$}}

\newcommand{\blzs}{\textsuperscript{$\blacklozenge$}}
\newcommand{\btrs}{\textsuperscript{$\blacktriangle$}}

\makeatletter
\newcommand\footnoteref[1]{\protected@xdef\@thefnmark{\ref{#1}}\@footnotemark}
\makeatother

\begin{document}

\title{
Enhancing Test Efficiency through Automated ATPG-Aware Lightweight Scan Instrumentation
}

\author{
\IEEEauthorblockN{Sudipta Paria, Md Rezoan Ferdous, Aritra Dasgupta, Atri Chatterjee and Swarup Bhunia\\
\IEEEauthorblockA{Department of Electrical and Computer Engineering, University of Florida, Gainesville, FL, USA\\
Email: \{sudiptaparia, mdrezoan.ferdous, aritradasgupta, a.chatterjee\}ufl.edu, swarup@ece.ufl.edu}
}
}

\maketitle
\thispagestyle{empty}
\pagestyle{empty}

\begin{abstract}

Scan-based Design-for-Testability (DFT) measures are prevalent in modern digital integrated circuits to achieve high test quality at low hardware cost. With the advent of 3D heterogeneous integration and chiplet-based systems, the role of scan is becoming ever more important due to its ability to make internal design nodes controllable and observable in a systematic and scalable manner. However, the effectiveness of scan-based DFT suffers from poor testability of internal nodes for complex circuits at deep logic levels. Existing solutions to address this problem primarily rely on Test Point Insertion (TPI) in the nodes with poor controllability or observability. However, TPI-based solutions, while an integral part of commercial practice, come at a high design and hardware cost. To address this issue, in this paper, we present \lite, a novel ATPG-aware lightweight scan instrumentation approach that utilizes the functional flip-flops in a scan chain to make multiple internal nodes observable and controllable in a low-cost, scalable manner. We provide both circuit-level design as well as an algorithmic approach for automating the insertion of \lite~for design modifications. We show that \lite~significantly improves the testability in terms of the number of patterns and test coverage for ATPG and random pattern testability, respectively, while incurring considerably lower overhead than TPI-based solutions.

\end{abstract}


\section{Introduction}
\label{sec:intro}

Over the last few decades, advances in fabrication techniques have made it possible to integrate more complex systems onto a single silicon chip. While this has brought many benefits, it has also introduced a significant challenge: testing these systems both after fabrication and during their use in the field. The problem lies in the growing complexity of the systems being tested, as well as the difficulty in accessing the internal parts of these systems. As digital designs became more complex, traditional testing methods that rely on accessing internal nodes became inefficient, making it difficult to observe and control the internal states of modern chips, which significantly impacts testability analysis. Testability of a design refers to the ability to efficiently and effectively test a circuit or system to ensure it performs as intended and to detect any faults or defects. Faults can manifest in different forms, such as stuck-at faults, and if left undetected, such faults can lead to operational malfunctions, increased power consumption, or even complete system failure. Scan chains improve testability by transforming the internal structure of a circuit into a more accessible form. A series of flip-flops (FFs) connected in linear sequence allows the circuit to be shifted into a known state, enabling individual control and observation of each flip-flop. This structure simplifies fault detection by providing easy access to internal nodes and ensuring systematic testing of all nodes in the circuit.

Recent research into testing complex systems has identified several factors contributing to the increasing testing costs. These factors include higher costs associated with test pattern generation and fault simulation, the growing volume of test data that must be stored and processed, and longer test application times. In response to these challenges, several strategies have been explored to mitigate the rising testing costs. These strategies include the development of more advanced test generation algorithms, the adoption of design for testability (DFT) techniques, and the implementation of built-in test methods. Built-in Self Test (BIST)\cite{bist} and Logic Built-in Self Test (LBIST) \cite{lbist} techniques emerged as popular DFT techniques that involve the use of on-chip test pattern generation and output response analysis to improve test time with higher coverage but suffers due to random-pattern-resistant (RPR) faults, which have low detection probabilities resulting in limited fault coverage with pseudo-random patterns. Test Point Insertion (TPI) \cite{tpi,tpi_1,tpi_2,tpi_3,tpi_4} has been studied extensively, which inserts extra gates as test points (TPs) into the circuit to improve the test coverage. Authors in \cite{tpi_touba} proposed a method that uses functional FFs to drive control test points instead of test-dedicated FFs for reducing the area overhead. However, TP insertion introduces significant hardware overhead due to the insertion of additional logic circuitry and also impacts critical path delay, resulting in limited performance for testability improvement. With the emergence of machine learning, various ML-based approaches \cite{ml_tpi_1,ml_tpi_2,ml_tpi_3,ml_tpi_4} were proposed for TP insertion for test coverage improvement, but they bounded by the heuristic algorithm and provide sub-optimal solutions due to inaccuracy arises from limited training and does not scale well with lack of generalization for larger designs. 

Several behavioral-level DFT approaches at the register transfer level (RTL) have been proposed to generate testable circuits for partial scan, sequential ATPG, and BIST, including modifications to RTL scan selection \cite{test_1,test_2}, improvements to synthesized circuits \cite{test_3}, and incorporating testability during behavioral synthesis \cite{test_4}. High-level techniques enhance datapath testability, assuming separate controller testing and full controllability of control signals in test mode \cite{test_5}. RTL testability analysis methods for sequential ATPG \cite{test_6} and Verilog models \cite{test_7} improve fault coverage, while VHDL testability analysis and test point insertion are discussed in \cite{vhdl_test}. Efforts to improve test pattern generation have been made at both the gate level and higher abstraction levels. Systems like PODEM-X \cite{podem} generate tests for large modules (up to 50,000 gates), while other advanced algorithms have been adapted for higher abstraction levels \cite{test_gen_1}, using functional blocks or behavioral descriptions. Commercial electronic design automation (EDA) tools can automate test pattern generation and fault simulation by integrating scan chains into the design, enabling efficient fault detection and comprehensive internal node testing. 

\begin{figure}[!htbp]
\centering
\subfloat[Traditional scan operation.]{\includegraphics[width=0.9\columnwidth]{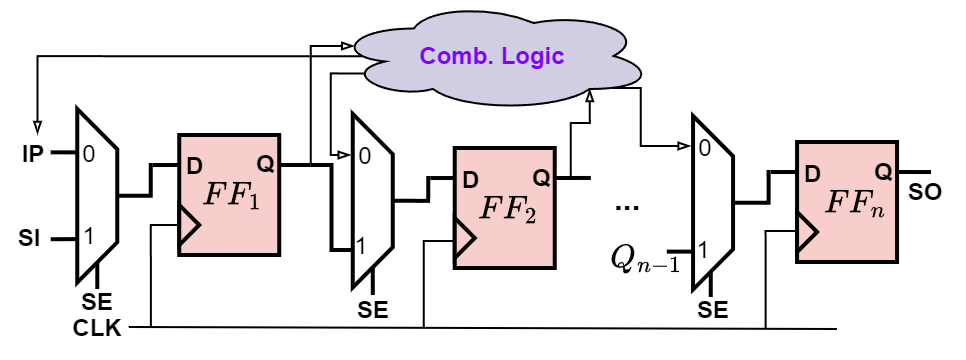}
\label{fig:trad_scan}}
\hfill
\subfloat[Scan operation with \lite.]{\includegraphics[width=1.0\columnwidth]{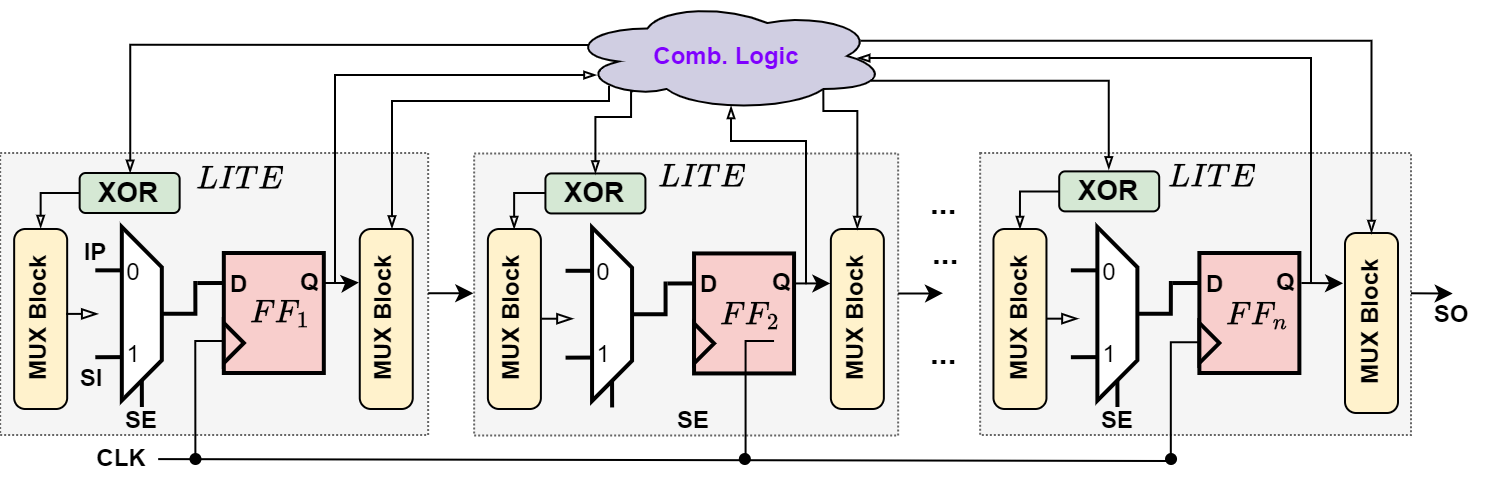}
\label{fig:lite_scan}}
\caption{Traditional scan operation and improved scan operation with \lite~integration in a design.}%
\label{fig:scan}%
\end{figure}

\begin{figure}[!htbp]%
    \centering
    \subfloat[][]{
    \includegraphics[scale=0.22]{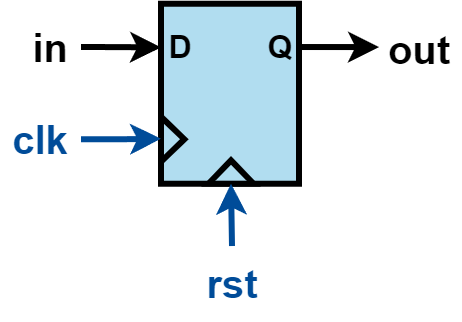}
    \label{fig:ff}}\hfill
    \subfloat[][]{
    \includegraphics[scale=0.16]{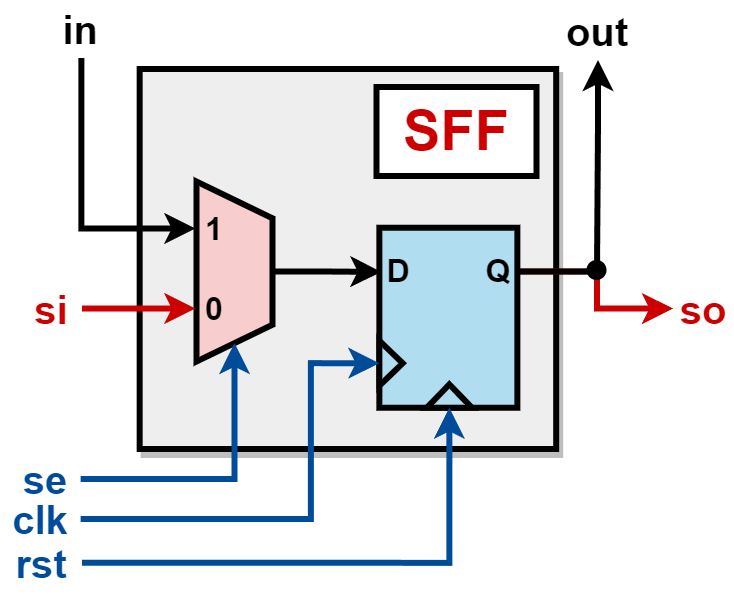}
     \label{fig:sff}}\hfill
    \subfloat[][]{
    \includegraphics[scale=0.16]{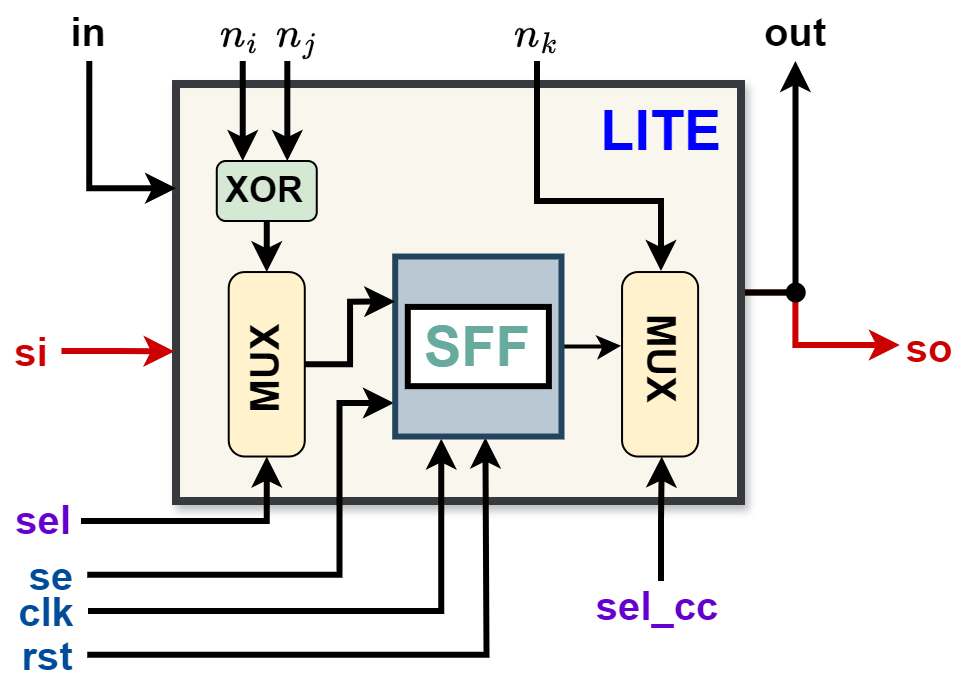}
     \label{fig:lite_sff}}
    \caption{Elements of a scan chain: (a) A single original FF with no scan. (b) Original FF driven by MUXed logic in traditional scan architecture, and (c) The single unit of the scan FF integrated with \lite.}
    \label{fig:ff_structure}
\end{figure}

Testability analysis can be limited due to the low controllability and observability of several nets in a circuit, which can hinder fault detection and reduce test coverage. TPI techniques often result in significant overhead, highlighting the need for more efficient DFT methods that minimize the area overhead along with the number of test patterns, thereby reducing test cycles and overall complexity, especially for larger circuits. In this paper, we propose \lite, a \textbu{L}ightweight Scan \textbu{I}nstrumentation for enhancing the post-silicon \textbu{T}est \textbu{E}fficiency in ICs. Fig. \ref{fig:trad_scan} demonstrates the traditional scan operation, and Fig. \ref{fig:lite_scan} illustrates the integration of \lite~into the existing circuit without altering the original functionality or changing the scan path for achieving improved testability. Fig. \ref{fig:ff_structure} represents the generic block diagrams of D flip-flop without scan (Fig. \ref{fig:ff}), traditional MUX-based scan flip-flop (SFF), as shown in Fig. \ref{fig:sff}, and the proposed \lite~incorporated scan flip-flop, as shown in Fig. \ref{fig:lite_sff}.

This paper makes the following major contributions:

\begin{itemize}
    \item It presents \lite, a circuit-level scan instrumentation technique that incorporates low-overhead additional logic to the scan flip-flops to improve the controllability and observability of internal nodes that are challenging to adequately cover by the scan infrastructure. \lite~eliminates the need for dedicated flip-flops to enable test point insertion. Instead, it inserts additional logic at the input (for enhancing observability) and output (for enhancing controllability) of a scan flip-flop to make a design more testable at low hardware overhead. The modified design is subsequently used by an automatic test pattern generation (ATPG) tool, which takes advantage of this extra hardware to generate patterns more efficiently, leading to reduced patterns and/or higher coverage. 
    \item It presents an algorithmic process for automatically modifying a scan chain without impacting the scan as well as normal mode design functionality. The additional logic uses standard cell library elements (e.g., MUX, XOR) only, and hence, \lite~works for any process technology and any library. Consequently, \lite~can be seamlessly integrated with commercial DFT tool flow.
    \item It provides an extensive evaluation of \lite~for ATPG patterns using a suite of ISCAS89 and ITC99 benchmarks. For ATPG patterns, we show that \lite~can improve the test pattern count by about 31\%.
    \item It provides analysis results for random patterns. We show that \lite~can effectively address the coverage issues arising from the random-resistant faults in a design, thereby improving the random pattern testability of these designs. Such a capability can be very attractive for built-in-self-test (BIST) solutions, including BIST for emerging 3D ICs, which rely on random test patterns for self-testing a logic circuit.
\end{itemize}

\vspace{-2em}

The rest of the paper is organized as follows: Section \ref{sec:back} describes relevant background and motivation behind this work; Section \ref{sec:method} presents the automated insertion of different \lite~configurations into a netlist; Section \ref{sec:results} demonstrates the experimental results and analysis on different open-source benchmarks showing the effectiveness of \lite~instrumentation over traditional ATPG; Section \ref{sec:discuss} discusses overhead improvements and potential extension of \lite~approach; and Section \ref{sec:conclude} concludes the paper.

\setlength\dashlinedash{1pt}
\setlength\dashlinegap{2pt}
\setlength\arrayrulewidth{0.5pt}
\def\arraystretch{1.25}
\begin{table*}[!htbp]
\centering
\caption{Comparison of the scope and features of \lite~with existing scan instrumentation approaches}
\label{tab:background}
\resizebox{0.95\textwidth}{!}{%
\begin{tabular}{ccccccc}
\hline
\textbf{Techniques} & \textbf{Scope} & \textbf{Overhead} & \textbf{Approach} & \textbf{Cov. Imprv.} & \textbf{Test Cycles Imprv.} & \textbf{Compatibility} \\ \hline
\begin{tabular}[c]{@{}c@{}} TPI Insertion \\ e.g. \cite{tpi,tpi_1,tpi_touba} \end{tabular}  & \begin{tabular}[c]{@{}c@{}}Improving Testability\end{tabular}  & Moderate & CC Improvement &  High & Medium &  ATPG    \\ \hdashline
\begin{tabular}[c]{@{}c@{}} Scan Reordering or Gating \\ e.g. \cite{scan_reorder_1},\cite{scan_reorder_2},\cite{scan_reorder_3} \end{tabular} & 
\begin{tabular}[c]{@{}c@{}}Improving Test Power\end{tabular}   & Moderate & Dynamic Reordering & N/A & N/A &  ATPG   \\ \hdashline
\begin{tabular}[c]{@{}c@{}} Scan Locking or Obfuscation \\ e.g. \cite{scan_locking_1},\cite{scan_locking_2}, \cite{bhunia2024invisible} \end{tabular} & 
\begin{tabular}[c]{@{}c@{}}Improving Scan Security\end{tabular}   & Low & Authenticate Scan Access & N/A & N/A &  ATPG   \\ \hdashline
\begin{tabular}[c]{@{}c@{}}\lite*\\(This work)\end{tabular}        & \begin{tabular}[c]{@{}c@{}}Improving Testability \\ Test Patterns or Coverage\end{tabular}          & Low      & \begin{tabular}[c]{@{}c@{}}Both CC and Obs \\Improvement\end{tabular} & High & High & ATPG, BIST \\ \hline
\end{tabular}%
}
\vspace{-1em}
\end{table*}

\section{Background}
\label{sec:back}

Scan-based DFT has proven to be highly effective for improving the structural test efficiency of a design. However, with the increasing complexity of modern designs, it faces a major challenge to test many internal nodes of a design that suffers from poor controllability and/or testability. To address this issue, test-point insertion is used as a DFT technique to make these nodes more testable during test application, leading to reduction in ATPG patterns count and/or improvement in test coverage. This is achieved by identifying hard-to-control or observe nodes in the design and intercepting with additional test point circuitry, which includes combinational logic gates and dedicated control/observe flip-flops. 

\subsection{SCOAP Values}

The Sandia Controllability and Observability (SCOAP) \cite{scoap} values help in determining how easily each net (or signal) in a circuit can be controlled and observed during testing, which are essential aspects of effective test pattern generation and fault detection. The SCOAP values can be computed for each net in the circuit, which provides a quantitative way of assessing how testable the nodes are. They help in guiding test point insertion, optimizing test pattern generation, improving fault coverage, and enhancing overall testability. 

\begin{itemize}
\item  \textit{Controllability (CC)} refers to the difficulty with which a signal or node in a circuit can be set to a desired value by applying test inputs. Nodes with low controllability values denote nodes that are hard-to-control to influence the value of 0 or 1. For example, for all primary inputs (PIs), the controllability values for setting-to-0 (${CC}\_{0}$) and setting-to-1 (${CC}\_{1}$) are both set to 1.

\item \textit{Observability (Obs)} refers to the ease with which the value of a signal or node can be observed or measured from the outputs of the circuit. Nodes with high observability (easy-to-observe) implies easy detection during testing. For example, for all primary outputs (POs), the observability value (${CC}\_{obs}$) is set to 0.
\end{itemize}

\subsection{Commercial EDA tools for Testability Analysis}

Commercial EDA tools (e.g., Synopsys TestMax, Cadence Modus, etc.) integrate advanced algorithms for ATPG, fault simulation, and scan chain insertion, enabling efficient detection of manufacturing defects and functional failures and providing automated solutions to improve the testability of digital circuits. These tools are efficient in performing testability analysis by evaluating the controllability and observability of circuit nodes, identifying areas of low fault coverage, and recommending or implementing test point insertion and scan design modifications. They can also generate optimized test patterns that minimize the number of required patterns while maximizing fault coverage. Furthermore, commercial tools are capable of analyzing circuits at different abstraction levels, from RTL to gate-level netlist, offering flexibility and scalability for complex designs. With their ability to simulate fault behavior, automate test generation, and provide valuable feedback on testability, commercial EDA tools significantly reduce manual effort and improve the overall efficiency of the test process, ensuring fault-tolerant integrated circuits. 

\subsection{Prevalent Challenges}

Testability analysis for larger circuits presents several challenges that impede effective testing and fault detection. One of the primary difficulties is the identification of an optimal number of low observability and low controllability nets, as well as the optimal placement of test points strategically to enhance observability and controllability. However, this process is not straightforward, as it requires balancing between maximizing fault coverage and minimizing the area overhead. The insertion of test points introduces significant area overhead, which can adversely affect the circuit's critical path, impacting performance and timing. Furthermore, the number of test point insertions can become excessive, leading to an increase in design size and complexity. Commercial EDA tools are equipped with automated test pattern generation capabilities but often struggle with incomplete fault analysis or reduced fault coverage and a significantly large number of undetected faults. The sheer volume of patterns needed for large designs leads to substantial test cycles, making the testing process time-consuming and resource-intensive. Additionally, commercial ATPG tools may fail to provide optimal test pattern sets, covering different fault models for complex circuits. Therefore, improving testability analysis with the reduction in the number of patterns with acceptable area overhead requires the integration of more scalable and robust DFT techniques that can be integrated with commercial EDA tools for achieving optimal performance for complex circuits.

\begin{figure*}[!ht]
\centering
\includegraphics[width=\textwidth]{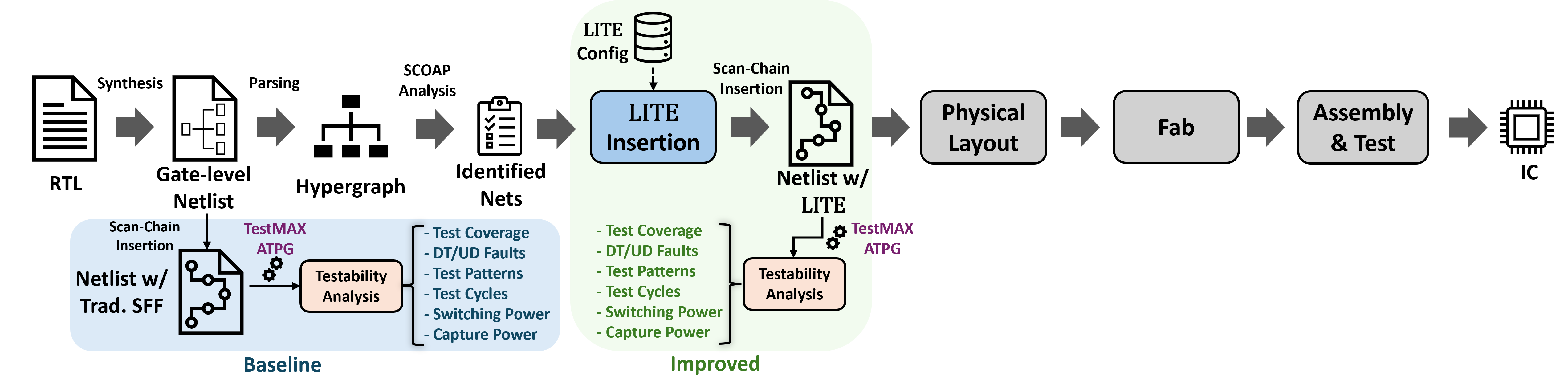}
\caption{Integrating \lite~into the existing ASIC design flow and interfacing with commercial DFT tool (e.g., Synopsys TestMax) for improved testability.}
\label{fig:lite_flow}
\vspace{-1em}
\end{figure*}

\subsection{Motivation}

Table \ref{tab:background} compares the \lite~approach with existing scan instrumentation approaches which vary in terms of scope and features. The primary goal of these approaches varies considerably, which includes testability (test coverage or pattern count), test power, and security improvement. Testability improvement solutions essentially rely on judicious insertion of controllable/observable test points using dedicated flip-flops or input/output ports. Some earlier work has attempted using a combination of dedicated and functional flip-flops, particularly for controllable TP insertion \cite{tpi_touba}. However, as discussed in Section II.C, existing TPI solutions come at high hardware overhead that impacts their scalability, may impact test time due to dedicated flip-flops in the scan chain, and do not typically consider improving random pattern testability. Hence, there is a critical need to augment the scan architecture that addresses the above challenges associated with TP insertion. The proposed \lite~approach addresses the above challenges by introducing a systematic approach to scan instrumentation to enhance both ATPG and random pattern testability at much lower hardware cost than TP insertion approaches.

\section{Proposed Methodology}
\label{sec:method}

In this section, we present the automated flow for insertion of \lite~into any gate-level netlist and illustrate different \lite~configurations for improving test efficiency.

Fig. \ref{fig:lite_flow} presents the integration of \lite~into the generic IC design flow and highlights the baseline and improved testability analysis with \lite. Algorithm \ref{algo:lite} demonstrates the sequence of steps involved for inserting \lite~into any given netlist ($\mathcal{I_N}$). Now we briefly describe the major stages of the proposed automated toolflow for incorporating \lite~into any RTL or gate-level design with the flexibility to select the preferred configuration ($\theta$) $\in$ \{\textbf{Config1\_Obs}, \textbf{Config2\_Obs}, \textbf{Config1\_Obs\_CC}, \textbf{Config2\_Obs\_CC}\} and generate the updated netlist with \lite, as discussed below.

\subsection{Hypergraph Generation and Topological Sort}

The automated toolflow generates hypergraph $\mathcal{G}$ for the given input gate-level netlist $\mathcal{I_N}$ with the graph represented as: $\mathcal{G} = (\mathcal{V},\mathcal{E})$ such that $\mathcal{V}$ denotes the set of gates in the design and $\mathcal{E}$ represents the interconnects between the gates. The proposed toolflow employs topological sorting of the hypergraph $\mathcal{G}$ for generating the linear ordering of vertices such that for each pair of vertices $\{u-v\}$ with a directed edge, $u$ must come before $v$ in the ordering. The topologically sorted graph $\mathcal{G_T}$ ensures that combinational loops are avoided and that the correct ordering of vertices is maintained.

\subsection{SCOAP Analysis}

The automated toolflow incorporates the calculation of Sandia Controllability and Observability (SCOAP) Values \cite{scoap} to identify the low controllable and low observable nets. Nets with low controllability values denote hard-to-control to the value of 0 or 1. Nodes with low observability values imply hard-to-observe detection during testing. The low controllable nets are appended to either $\mathcal{CC}_0$ or $\mathcal{CC}_1$ based on the hard controllability value of 0 or 1, respectively, while low observable nets are appended to $\mathcal{CC}_{obs}$ for further analysis in the subsequent stages. 

\vspace{-1em}
\subsection{\lite~Insertion}

\noindent $\bullet$ \textbf{\lite: Config1\_Obs}

\begin{figure}[!htbp]
\centering
\subfloat[\lite: Config1\_Obs.]{\includegraphics[width=0.5\columnwidth]{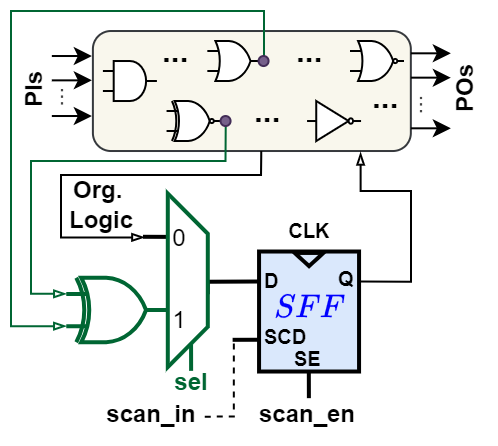}
\label{fig:lite_config1}}
\subfloat[\lite: Config2\_Obs.]{\includegraphics[width=0.5\columnwidth]{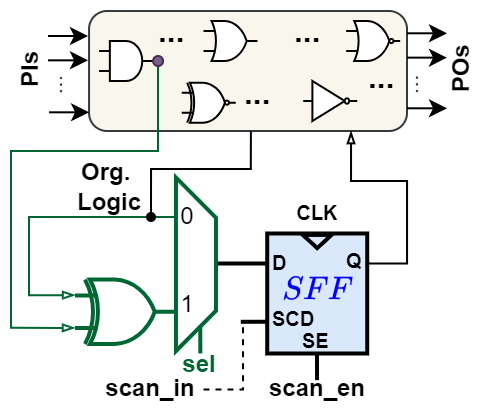}
\label{fig:lite_config2}}
\caption{\lite: Configuration with observability improvement logic (no CC).}%
\label{fig:lite_config}%
\end{figure}

This configuration selects two less observable nets ($obs_1$ and $obs_2$) based on SCOAP values from the sorted list $\mathcal{CC}_{obs}$ such that the topological order is satisfied and the conflict-PI analysis yields \textit{True} denoting these two nets are non-conflicting without contradictory values for PIs. The selected nets are routed to an XOR$2$ gate, followed by a $2\times1$ MUX. The original logic, previously connected to the D input of the scan FF, is now connected to the MUX\_0 input. The output of the XOR gate is routed to the MUX\_1 input. Each MUX is controlled by a common select (\textit{sel}) line configured as an additional PI. Fig. \ref{fig:lite_config1} illustrates the circuit diagram of Config1\_Obs configuration with the additional logic for \lite~marked in \textcolor{ForestGreen}{green}.

\noindent $\bullet$ \textbf{\lite: Config2\_Obs}

This configuration selects one less observable net ($obs_1$) based on SCOAP values from the sorted list $\mathcal{CC}_{obs}$ such that the topological order is satisfied. The selected net is then routed to an XOR$2$ gate with other input connected to the original logic, previously connected to the D input of the scan FF. The XOR gate is followed by a $2\times1$ MUX with original logic and XOR output connected to the MUX\_0 and MUX\_1 inputs, respectively. The select line (\textit{sel}) is set to value $1$. Fig. \ref{fig:lite_config2} shows the circuit diagram of Config2\_Obs configuration with the additional logic for \lite~marked in \textcolor{ForestGreen}{green}.

\begin{figure}[!htbp]
\centering
\subfloat[\lite: Config1\_Obs\_CC.]{\includegraphics[width=0.5\columnwidth]{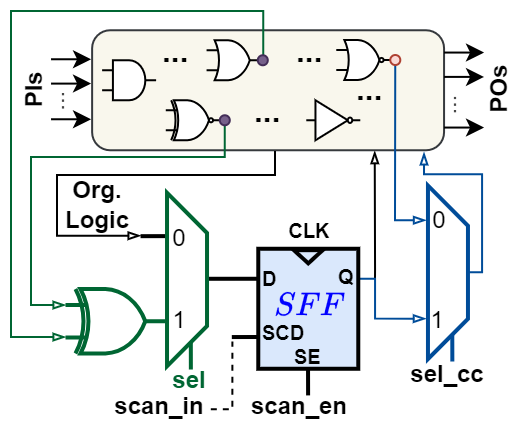}
\label{fig:lite_config1_cc}}
\subfloat[\lite: Config2\_Obs\_CC.]{\includegraphics[width=0.5\columnwidth]{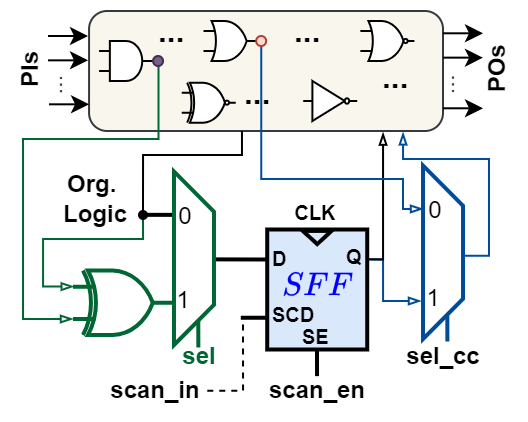}
\label{fig:lite_config2_cc}}
\caption{\lite: Configuration with observability (Obs.) and controllability (CC) improvement logic.}%
\label{fig:lite_config_cc}%
\vspace{-1em}
\end{figure}

\noindent $\bullet$ \textbf{\lite: Config1\_Obs\_CC}

To improve the testability further, we consider the low controllable nets to values 0 or 1 from the sorted lists $\mathcal{CC}_{0}$ and $\mathcal{CC}_{1}$, respectively. The circuit architecture of \textit{Config1\_Obs} is extended with an additional $2\times1$ MUX at the Q output of the scan FF, with the same connected to MUX\_0 and the one of the low controllable selected nodes to MUX\_1 port. For each insertion, one low controllable node is selected from either $\mathcal{CC}_{0}$ or $\mathcal{CC}_{1}$ alternatively. Each MUX is controlled by a common select (\textit{sel\_cc}) line configured as an additional PI. Fig. \ref{fig:lite_config1_cc} depicts the circuit diagram of Config1\_Obs\_CC configuration with Config1\_Obs marked in \textcolor{ForestGreen}{green} and the additional logic for controllability marked in \textcolor{RoyalBlue}{blue}.

\noindent $\bullet$ \textbf{\lite: Config2\_Obs\_CC}

Similar to Config1\_Obs\_CC, we also extend the \textit{Config2\_Obs} to improve testability by including low controllable nets selected alternatively from $\mathcal{CC}_{0}$ and $\mathcal{CC}_{1}$, respectively. The additional $2\times1$ MUX is placed in such a way that MUX\_0 is connected to the Q output of the scan FF, while MUX\_1 port is connected to the selected low controllable net driven by a common select (\textit{sel\_cc}) line configured as an additional PI. Fig. \ref{fig:lite_config2_cc} presents the circuit diagram of Config2\_Obs\_CC configuration with Config2\_Obs marked in \textcolor{ForestGreen}{green} and the additional logic for controllability marked in \textcolor{RoyalBlue}{blue}. \\

\noindent $\bullet$ \textbf{Rationale for Selecting XOR and MUX in \lite}

By introducing XOR gates, low-observable and low-controllable points in the circuit can be dynamically manipulated by ATPG to achieve desired values, thereby improving the controllability and observability of such hard-to-observe nets. XOR gate exhibits a unique equiprobable masking effect when its inputs are randomly distributed. This property ensures that the XOR gate does not bias the output towards a specific value, unlike AND and OR gates. XOR gates help improve fault coverage by providing an additional degree of freedom in controlling the values of low observable nodes during pattern generation. The addition of MUX allows the original logic to be combined with the newly controlled logic. This method ensures that various states of the circuit can be accessed and monitored more easily, even for previously difficult-to-test points. The XOR operation enables the injection of specific signal variations without altering the original functionality, and the MUX allows for switching between the original and modified logic paths. By leveraging this combination, it becomes possible to more effectively target and detect faults across the circuit, which leads to the generation of more efficient and compact test patterns. Alternatively, we can employ XNOR gate instead of XOR since both of them exhibit unique equiprobable masking effects without any bias.

\vspace{-1em}
\subsection{\lite-inserted Netlist Generation}

The automated toolflow translates the modified hypergraph after insertion of \lite~at respective positions for the given configuration ($\theta$) into gate-level netlist $\mathcal{S_N}$ by analyzing and extracting the necessary connectivity information and arrangement of gates and their interconnections. The generated netlist can be directly used by EDA tools for improved testability analysis and further stages of IC design flow.

\begin{algorithm}[!htbp]
\caption{\lite~insertion}
\label{algo:lite}
\DontPrintSemicolon 
  \SetKwInOut{Input}{Input}
  \SetKwInOut{Output}{Output}
  \Input{Netlist ($\mathcal{I_N}$), Config ($\theta$)}
  \Output{Updated netlist with \lite~($\mathcal{S_N}$)}

  $\mathcal{G}$ $\leftarrow$ \text{parse $\mathcal{I_N}$ and construct hypergraph $\mathcal{G} = \{\mathcal{V},\mathcal{E}\}$}\\
  $\mathcal{G^T}$ $\leftarrow$ \text{\textbf{topological\_sort}($\mathcal{G}$)}\\
  $\mathcal{CC}_0,\mathcal{CC}_1,\mathcal{CC}_{obs}$ $\leftarrow$ \textbf{scoap\_analysis}($\mathcal{G},\mathcal{G_T}$) \\
  \If{$\theta$ is config$1$\_Obs}
  {
        $obs_1,obs_2 \leftarrow$ \textbf{select\_nets\_obs($\mathcal{CC}_{obs}$)} \\
        \If{\textbf{check\_topo\_order($\mathcal{G_T},obs_1,obs_2$)} and \textbf{check\_conflict\_pi($\mathcal{G},obs_1,obs_2$)}}
        {  
           $\mathcal{G} \leftarrow$ \textbf{insert\_\lite}($\mathcal{G},obs_1,obs_2,\theta$)  
        }   
  }
  \ElseIf{$\theta$ is config$2$\_Obs}
  {
        $obs_1 \leftarrow$ \textbf{select\_nets\_obs($\mathcal{CC}_{obs}$)} \\
        \If{\textbf{check\_topo\_order($\mathcal{G_T},obs_1$)}}
        {  
           $\mathcal{G} \leftarrow$ \textbf{insert\_\lite}($\mathcal{G},obs_1,\theta$)  
        }   
  }
  \If{$\theta$ is config$1$\_Obs\_$CC$ or config$2$\_Obs\_$CC$}
  {
        $cc_0,cc_1 \leftarrow$ \textbf{select\_nets\_cc($\mathcal{CC}_0,\mathcal{CC}_1$)} \\
        \If{\textbf{check\_topo\_order($\mathcal{G_T},cc_0,cc_1,FF_{out}$)}}
        {  
           $\mathcal{G} \leftarrow$ \textbf{insert\_\lite}($\mathcal{G},cc_0,cc_1,FF_{out},\theta$)  
        }   
  }
  $\mathcal{S_N} \leftarrow$ \textbf{write\_verilog\_from\_graph}($\mathcal{G}$) \\
  \textbf{return} $\mathcal{S_N}$
\end{algorithm}

\vspace{-2em}

\section{Results and Analysis}
\label{sec:results}

In this section, we present our experimental setup and describe the results obtained on multiple open-source benchmarks that demonstrate the robustness and efficacy of the proposed \lite~instrumentation for testability and fault coverage. We also compare the PPA (Power, Performance, Area) overhead values, fault coverage, and testability metrics observed from different configurations of our proposed approach against the traditional ATPG-based DFT setup as a baseline.
\def\arraystretch{1.15}
\newcommand{\iscas}{\multirow{5}{*}{ISCAS89\footref{iscas}}}
\newcommand{\itc}{\multirow{5}{*}{ITC99\footref{itc}}}
\begin{table*}[!htbp]
\centering
\caption{Design specifications and fault coverage metrics for the evaluation benchmarks.}
\label{tab:benchmarks}
\resizebox{\textwidth}{!}{%
\begin{tabular}{c|c|cccccc|cccccc}
\hline
\multirow{2}{*}{\textbf{Design}} &
\multirow{2}{*}{\textbf{Source}} &
\multicolumn{6}{c|}{\textbf{Gate-Level Netlist}\blzs} &
\multicolumn{6}{c}{\textbf{Fault Coverage and Testability Analysis (Baseline ATPG)}} \\ \cline{3-10} \cline{11-14}
 & & \textbf{\#PIs} &
\textbf{\#POs} &
\textbf{\#FFs} &
\textbf{\#Gates} &
\textbf{Area ($\mu$$m^{2}$)} &
\textbf{Delay ($ns$)} &
\textbf{DT}\bstars &
\textbf{Cov\%}\bstars &
\textbf{\#Patterns} &
\textbf{\#Cycles} &
\textbf{\textbf{ASS\%}\bstars} &
\textbf{\textbf{ACS\%}\bstars} \\ \hline
s9234   & \iscas & 42 & 40  & 145  & 729   & 6880.35   & 2.14 & 4654   & 100\% & 186   & 27671    & 49.40\% & 34.36\% \\
s13207  &        & 68 & 153 & 625  & 2315  & 24850.08  & 3.46 & 14172  & 100\% & 453   & 284939   & 48.66\% & 24.31\% \\
s15850  &        & 83 & 151 & 513  & 2721  & 23403.70  & 4.87 & 17050  & 100\% & 587   & 303385   & 48.31\% & 30.55\% \\
s38584  &        & 44 & 305 & 1275 & 7950  & 70654.01  & 3.59 & 51997  & 100\% & 837   & 1070811  & 48.27\% & 24.32\% \\
s38417  &        & 32 & 107 & 1564 & 7617  & 65455.28  & 2.29 & 48178  & 100\% & 1198  & 1878777  & 47.59\% & 26.30\% \\ \hdashline
b17     & \itc   & 41 & 97  & 1322 & 11700 & 104707.92 & 8.77 & 87864  & 100\% & 2950  & 3910071  & 46.65\% & 10.63\% \\
b18     &        & 40 & 24  & 3030 & 34458 & 325823.74 & 8.64 & 250544 & 100\% & 6309  & 19138167 & 46.93\% & 10.90\% \\
b19     &        & 49 & 31  & 6062 & 63187 & 583205.58 & 8.25 & 464592 & 100\% & 11734 & 71172637 & 47.61\% & 9.51\% \\
b20     &        & 36 & 23  & 430  & 4838  & 44398.83  & 8.78 & 35083  & 100\% & 772   & 334705   & 47.07\% & 15.58\% \\
b22     &        & 36 & 23  & 613  & 7337  & 64977.32  & 7.78 & 53082  & 100\% & 1101  & 678827   & 47.68\% & 14.71\% \\ \hline
\textbf{Average} &  & 
\textbf{47} & \textbf{95} & \textbf{1558} & \textbf{14285}  & \textbf{131435.68} & \textbf{5.86} &
\textbf{102722}  & \textbf{100\%} & \textbf{2613} & \textbf{9879999} & \textbf{47.82\%} & \textbf{20.12\%} \\ \hline
\end{tabular}%
}
\scriptsize
\raggedright \\
\blzs Values obtained for gate-level netlist (mapped to SkyWater 130nm\footref{skywater}); reported by the synthesis tool.\\
\bstars \textbf{DT:} Detected stuck-at faults; \textbf{Cov\%:} Percent of DT covered; \textbf{ASS:} Average Shift Switching; \textbf{ACS:} Average Capture Switching;
reported by the ATPG tool.
\end{table*}

\begin{table*}[!htbp]
\centering
\caption{Testability (\# Patterns and \# Cycles) improvements under different configurations in \lite.}
\label{tab:improve_pattern_cycle}
\resizebox{\textwidth}{!}{%
\begin{tabular}{c|ccc|ccc|ccc|ccc}
\hline
\multirow{3}{*}{\textbf{Design}} &
\multicolumn{3}{c|}{\textbf{Config1\_Obs}} &
\multicolumn{3}{c|}{\textbf{Config2\_Obs}} &
\multicolumn{3}{c|}{\textbf{Config1\_Obs\_CC}} &
\multicolumn{3}{c}{\textbf{Config2\_Obs\_CC}} \\ \cline{2-4} \cline{5-7} \cline{8-10} \cline{11-13}
 & 
\multirow{2}{*}{\textbf{DT~(Cov\%)}\bstars} & \multicolumn{2}{c|}{\textbf{Improv. (\%)}\btrs} &
\multirow{2}{*}{\textbf{DT~(Cov\%)}\bstars} & \multicolumn{2}{c|}{\textbf{Improv. (\%)}\btrs} &
\multirow{2}{*}{\textbf{DT~(Cov\%)}\bstars} & \multicolumn{2}{c|}{\textbf{Improv. (\%)}\btrs} &
\multirow{2}{*}{\textbf{DT~(Cov\%)}\bstars} &  \multicolumn{2}{c}{\textbf{Improv. (\%)}\btrs} \\ \cline{3-4} \cline{6-7} \cline{9-10} \cline{12-13}
 & 
 &
\textbf{\#Patterns} &
\textbf{\#Cycles} &
 &
\textbf{\#Patterns} &
\textbf{\#Cycles} &
 &
\textbf{\#Patterns} &
\textbf{\#Cycles} &
 &
\textbf{\#Patterns} &
\textbf{\#Cycles} \\ \hline
s9234   & 6762   (100\%) & 8.06\%  & 8.03\%  & 6337   (100\%) & 16.13\% & 16.06\% & 7900   (100\%) & 11.83\% & 11.77\% & 7456   (100\%) & 14.52\% & 14.46\% \\
s13207  & 23092  (100\%) & 24.72\% & 24.63\% & 21333  (100\%) & 40.84\% & 40.72\% & 28065  (100\%) & 46.58\% & 46.45\% & 26185  (100\%) & 48.12\% & 47.99\% \\
s15850  & 24516  (100\%) & 37.82\% & 37.75\% & 23033  (100\%) & 39.35\% & 39.28\% & 28591  (100\%) & 44.46\% & 44.39\% & 26906  (100\%) & 37.14\% & 37.07\% \\
s38584  & 69958  (100\%) & 17.08\% & 17.06\% & 66619  (100\%) & 23.42\% & 23.38\% & 80104  (100\%) & 44.56\% & 44.51\% & 76645  (100\%) & 45.52\% & 45.47\% \\
s38417  & 70002  (100\%) & 55.01\% & 54.96\% & 65409  (100\%) & 53.34\% & 53.30\% & 82273  (100\%) & 67.20\% & 67.14\% & 76822  (100\%) & 53.17\% & 53.13\% \\ \hdashline
b17     & 106228 (100\%) & 13.56\% & 13.55\% & 102406 (100\%) & 18.98\% & 18.98\% & 116775 (100\%) & 8.78\%  & 8.78\%  & 102406 (100\%) & 18.98\% & 18.98\% \\
b18     & 292205 (100\%) & 1.95\%  & 1.95\%  & 283892 (100\%) & 2.68\%  & 2.68\%  & 316352 (100\%) & 10.68\% & 10.68\% & 307881 (100\%) & 10.14\% & 10.14\% \\
b19     & 548726 (100\%) & 13.99\% & 13.99\% & 531270 (100\%) & 11.14\% & 11.14\% & 597117 (100\%) & 19.97\% & 19.97\% & 579267 (100\%) & 22.40\% & 22.40\% \\
b20     & 41069  (100\%) & 33.42\% & 33.38\% & 39817  (100\%) & 34.84\% & 34.80\% & 44466  (100\%) & 30.57\% & 30.53\% & 39817  (100\%) & 34.84\% & 34.80\% \\
b22     & 61611  (100\%) & 33.42\% & 33.39\% & 59829  (100\%) & 32.79\% & 32.76\% & 66423  (100\%) & 33.61\% & 33.58\% & 59829  (100\%) & 32.79\% & 32.76\% \\ \hline
\textbf{Average} &
\textbf{124417 (100\%)} & \textbf{\textbf{23.90\%}} & \textbf{23.87\%} &
\textbf{119995 (100\%)} & \textbf{\textbf{27.35\%}} & \textbf{27.31\%} &
\textbf{136807 (100\%)} & \textbf{\textbf{31.82\%}} & \textbf{31.78\%} &
\textbf{130321 (100\%)} & \textbf{\textbf{31.76\%}} & \textbf{31.72\%} \\ \hline
\end{tabular}%
}
\scriptsize
\raggedright \\
\btrs Percentage improvements in \#Patterns and \#Cycles compared to Baseline ATPG.\\
\bstars \textbf{DT~(Cov\%):} Detected stuck-at faults and the corresponding test coverage; reported by the ATPG tool. The higher fault counts correspond to the combinational logic (XOR/MUX) added by \lite, and are proportional to the \#FFs in the scan chain.
\vspace{-1em}
\end{table*}

\subsection{Experimental Setup}

\lite~was evaluated using multiple open-source IPs from the ISCAS89\footnote{https://ddd.fit.cvut.cz/www/prj/Benchmarks/\label{iscas}} and ITC99\footnote{https://github.com/cad-polito-it/I99T\label{itc}} benchmark suites. Synopsys Design Compiler (V-2023.12-SP5) was used to map the input RTL or gate-level netlist to the Skywater 130nm\footnote{https://github.com/google/skywater-pdk\label{skywater}} standard cell library and insert the DFT scan chain. To calculate the PPA metrics and overhead, all synthesis steps were constrained to an operating clock frequency of 100 MHz. Synopsys VCS (V-2023.12) was used to simulate the design functionality during the rareness (SCOAP) and coverage analysis steps. Synopsys TestMAX (V-2023.12-SP5) was used to generate the test patterns for fault coverage via ATPG. All of our experiments were performed on a Red Hat Enterprise Linux Server 7.9 operating system with an AMD® Epyc 7713 64-core processor and 1007.6 GiB memory.

\subsection{Evaluation Benchmarks}

The gate-level design specifications reported by the synthesis tool for the evaluation benchmarks, along with the corresponding fault coverage and testability analysis metrics reported by the ATPG tool, are listed in Table \ref{tab:benchmarks}. The gate/FF counts, area, and delay values are representative of the nominal variants of the standard cells from SkyWater 130nm and not the NAND2 gate-equivalent values. We choose five large benchmarks each from ISCAS89 and ITC99, with a wide range of functionalities to ensure that our analysis is fair and comprehensive. On average, the evaluation benchmarks have $\approx$ 14.23 K (thousand) combination gates and $\approx$ 1.56 K sequential FF, with 47 PIs and 95 POs. The DFT scan chain is inserted by the synthesis tool, and the resulting netlists have an area of $\approx$ 131.44 K ($\mu$$m^2$) with a delay of 5.86 $ns$ on average, which satisfies the synthesis constraints. The number of detected stuck-at faults (DT) varies from 4.65 K (s9234) up to 464.59 K (b19), with 100\% test coverage for each benchmark, for which the ATPG tool generated between 186 (s9234) and 11.73 K (b19) directed test patterns. The total test cycle count varies from 276.71 K for s9234 to 71.17 M (million) for b19. The average switching values reported by the ATPG tool for shifting (ASS) and capture (ACS) are 47.82\% and 20.12\%, respectively, when averaged over all evaluation benchmarks and can be used to estimate the switching power generated during the post-silicon validation at a testing facility.

\begin{table*}[!htbp]
\centering
\caption{Gate-level hardware overheads and average switching results under different configurations in \lite.}
\label{tab:ppa_overheads}
\resizebox{\textwidth}{!}{%
\begin{tabular}{c|cccc|cccc|cccc|cccc}
\hline
\multirow{3}{*}{\textbf{Design}} &
\multicolumn{4}{c|}{\textbf{Config1\_Obs}} &
\multicolumn{4}{c|}{\textbf{Config2\_Obs}} &
\multicolumn{4}{c|}{\textbf{Config1\_Obs\_CC}} &
\multicolumn{4}{c}{\textbf{Config2\_Obs\_CC}} \\ \cline{2-5} \cline{6-9} \cline{10-13} \cline{14-17}
 & 
\multicolumn{2}{c}{\textbf{Overhead (x)}\blzs} & \multirow{2}{*}{\textbf{ASS\%}\bstars} & \multirow{2}{*}{\textbf{ACS\%}\bstars} &
\multicolumn{2}{c}{\textbf{Overhead (x)}\blzs} & \multirow{2}{*}{\textbf{ASS\%}\bstars} & \multirow{2}{*}{\textbf{ACS\%}\bstars} &
\multicolumn{2}{c}{\textbf{Overhead (x)}\blzs} & \multirow{2}{*}{\textbf{ASS\%}\bstars} & \multirow{2}{*}{\textbf{ACS\%}\bstars} &
\multicolumn{2}{c}{\textbf{Overhead (x)}\blzs} & \multirow{2}{*}{\textbf{ASS\%}\bstars} & \multirow{2}{*}{\textbf{ACS\%}\bstars} \\ \cline{2-3} \cline{6-7} \cline{10-11} \cline{14-15}
 & 
\textbf{Area} &
\textbf{Delay} &
 &
 &
\textbf{Area} &
\textbf{Delay} &
 &
 &
\textbf{Area} &
\textbf{Delay} &
 &
 &
\textbf{Area} &
\textbf{Delay}
& & \\ \hline
s9234   & 1.53x & 1.10x & 49.02\% & 40.71\% &  1.53x & 1.05x & 49.14\% & 51.49\% & 1.77x & 2.02x & 48.98\% & 35.12\% &  1.77x & 1.96x & 49.40\% & 49.46\% \\
s13207  & 1.60x & 0.82x & 46.66\% & 42.89\% &  1.60x & 0.80x & 48.74\% & 48.39\% & 1.89x & 1.91x & 47.54\% & 42.51\% &  1.89x & 1.86x & 49.33\% & 47.03\% \\
s15850  & 1.59x & 0.66x & 47.08\% & 41.24\% &  1.59x & 0.64x & 49.46\% & 50.32\% & 1.84x & 1.54x & 47.55\% & 41.33\% &  1.84x & 1.55x & 49.76\% & 50.21\% \\
s38584  & 1.57x & 0.88x & 47.02\% & 36.61\% &  1.57x & 0.57x & 49.62\% & 46.68\% & 1.82x & 1.58x & 47.68\% & 37.46\% &  1.82x & 1.50x & 49.82\% & 41.38\% \\
s38417  & 1.50x & 1.05x & 47.68\% & 36.08\% &  1.50x & 1.04x & 49.79\% & 46.07\% & 1.72x & 2.85x & 47.79\% & 37.31\% &  1.72x & 2.85x & 49.77\% & 43.98\% \\ \hdashline
b17     & 1.26x & 1.33x & 46.02\% & 19.76\% &  1.26x & 1.35x & 49.56\% & 54.78\% & 1.40x & 1.74x & 42.70\% & 30.67\% &  1.26x & 1.35x & 49.56\% & 54.78\% \\
b18     & 1.06x & 2.11x & 46.75\% & 14.62\% &  1.06x & 1.95x & 50.12\% & 47.82\% & 1.17x & 2.73x & 41.74\% & 29.05\% &  1.17x & 2.70x & 50.51\% & 48.33\% \\
b19     & 1.14x & 1.88x & 47.19\% & 18.60\% &  1.14x & 1.83x & 49.86\% & 52.84\% & 1.25x & 2.74x & 43.70\% & 30.18\% &  1.25x & 2.74x & 49.73\% & 51.65\% \\
b20     & 1.09x & 2.24x & 47.09\% & 24.69\% &  1.09x & 2.05x & 49.43\% & 48.51\% & 1.19x & 2.46x & 48.12\% & 22.45\% &  1.09x & 2.05x & 49.43\% & 48.51\% \\
b22     & 1.10x & 2.52x & 47.25\% & 22.24\% &  1.10x & 2.32x & 49.42\% & 47.13\% & 1.20x & 2.70x & 48.29\% & 21.22\% &  1.10x & 2.32x & 49.42\% & 47.13\% \\ \hline
\textbf{Average} &
\textbf{1.34x} & \textbf{1.46x} & \textbf{47.18\%} & \textbf{29.74\%} &
\textbf{1.34x} & \textbf{1.36x} & \textbf{49.51\%} & \textbf{49.40\%} &
\textbf{1.53x} & \textbf{2.23x} & \textbf{46.41\%} & \textbf{32.73\%} &
\textbf{1.49x} & \textbf{2.09x} & \textbf{49.67\%} & \textbf{48.25\%}  \\ \hline
\end{tabular}%
}
\scriptsize
\raggedright \\
\blzs Overheads calculated as x times the original value; reported by the synthesis tool. \\
\bstars Average Shift Switching (ASS) and Average Capture Switching (ACS) as the percentage of total DFT scan length; reported by the ATPG tool.
\vspace{-1em}
\end{table*}

\subsection{\lite~vs. Baseline ATPG}

\subsubsection{Comparing Fault Coverage and Testability}

Table \ref{tab:improve_pattern_cycle} presents the fault coverage and testability metrics reported by the ATPG tool for 4 different configurations in \lite: (1) \textbf{Config1\_Obs}, (2) \textbf{Config2\_Obs}, (3) \textbf{Config1\_Obs\_CC} and (4) \textbf{Config2\_Obs\_CC}. For each configuration, the total count of detected stuck-at faults (DT) increases compared to the baseline due to the combinational logic (XOR and MUX) introduced by \lite~to the DFT scan chain and is directly proportional to the \#FFs present in the scan chain. Moreover, test coverage (Cov) remains at 100\% for each evaluation benchmark across all 4 configurations, implying that all faults, including the new ones added due \lite, are successfully covered by the test patterns generated by the ATPG. 

From Table \ref{tab:improve_pattern_cycle}, it can be seen that for all configurations of \lite, the number of test patterns (\#Patterns) and the total number of test cycles required (\#Cycles) both improve (reduce) significantly for each evaluation benchmark, resulting in better testability across the board. On average, \textbf{Config2\_Obs} provides better observability (and hence better testability) than \textbf{Config1\_Obs}, which is reflected by the improvement (reduction) in both \#Patterns and \#Cycles by equivalent margins. The additional controllability in \textbf{Config1\_Obs\_CC} and \textbf{Config2\_Obs\_CC} improves the testability further, and as a result, both \#Patterns and \#Cycles improves (reduces), as observed from their respective columns in Table \ref{tab:improve_pattern_cycle}. Interestingly, the difference in testability improvement between \textbf{Config1\_Obs} and \textbf{Config2\_Obs} goes away, resulting in equivalent \#Patterns and \#Cycles between \textbf{Config1\_Obs\_CC} and \textbf{Config2\_Obs\_CC}, with the former providing the best overall improvement in both \#Patterns and \#Cycles, which reduce by the average margins of 31.82\% and 31.78\%, respectively, compared to the baseline ATPG setup.

\subsubsection{Overhead Analysis}

Table \ref{tab:ppa_overheads} presents the hardware overheads for the 4 different configurations in \lite, calculated using the gate-level area and delay values reported by the synthesis tool. The overheads are expected due to the additional logic used for the scan modifications in \lite, but are within acceptable margins. Moreover, the area overheads incurred by \lite~are much lower than the traditional approach to improve testability, as discussed in the next section. Since the scan modifications for \textbf{Config1\_Obs} and \textbf{Config2\_Obs} are implemented using an identical set of logic gates, the corresponding area overheads are the same. Similarly, the area overheads for \textbf{Config1\_Obs\_CC} and \textbf{Config2\_Obs\_CC} are also identical for the ISCAS89 benchmarks, and those for ITC99 are also very close to each other, with minor variations due to timing constraints.
The average switching values for scan shifting (ASS) and scan capture (ACS), reported by the ATPG tool as a fraction (\%) of the total length DFT scan chain (\#FFs), are directly proportional to the switching power consumption during testing and can be used to estimate the overall test power. We observe that the scan chain modifications introduced by \lite~have a negligible impact on the cumulative value of ASS\% across all configurations, which is expected since the length of the scan chain (and hence the number of shifts required per pattern) remains the same. However, ACS\% increases significantly compared to baseline ATPG values, with the maximum capture switching observed for \textbf{Config2\_Obs}. The increase in ACS\% can be explained by the presence of the XOR gates added to the input of each SFF in \lite, as the XOR gate switches with a probability of 50\% for random inputs. In particular, the scan modifications in \textbf{Config2\_Obs} and \textbf{Config2\_Obs\_CC}, as seen in Fig. \ref{fig:lite_config2} and Fig. \ref{fig:lite_config2_cc} respectively, are more likely to switch during scan capture as the original logic is activated more frequently.

\begin{figure}[!htbp]
\centering
\subfloat[Control 0 Test Point Insertion.]{\includegraphics[width=0.8\columnwidth]{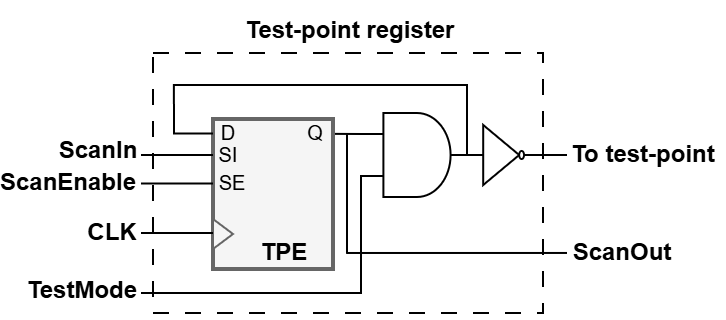}
\label{fig:TP_cc0}}
\hfill
\subfloat[Control 1 Test Point Insertion.]{\includegraphics[width=0.8\columnwidth]{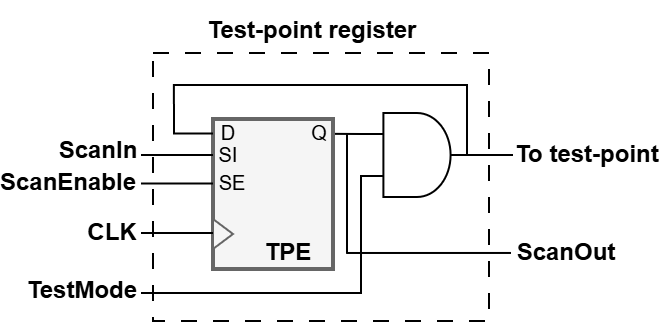}
\label{fig:TP_cc1}}
\hfill
\subfloat[Observable Test Point Insertion.]{\includegraphics[width=0.8\columnwidth]{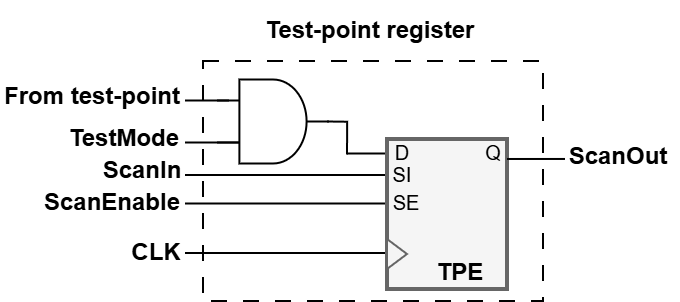}
\label{fig:TP_co}}
\caption{Test Point Insertion for hard-to-control and hard-to-observe points.}%
\label{fig:TPI_sota}%
\end{figure}

\begin{table}[ht]
\centering
\resizebox{\columnwidth}{!}{%
\begin{minipage}{0.06\columnwidth}
    \centering
    \begin{tabular}{|>{\columncolor[HTML]{D9EAD3}}l|c|}
    \hline
    \rowcolor{gray!30} 
    \textbf{Gate/FF} & \textbf{Area ($\mu$$m^2$)} \\
    \hline
    \rowcolor{gray!10} 
    DFF & 20.02 \\
    \rowcolor{gray!5} 
    SDFF & 26.28 \\
    \rowcolor{gray!10} 
    INV2 & 3.76 \\
    \rowcolor{gray!10} 
    AND2 & 6.26 \\
    \rowcolor{gray!5} 
    MUX2 & 11.26 \\
    \rowcolor{gray!10} 
    XOR2 & 8.76 \\
    \hline
    \end{tabular}
\end{minipage}%
\hspace{3.2cm} 
\begin{minipage}{0.75\columnwidth}
    \centering
    \begin{tabular}{|>{\columncolor[HTML]{D9EAD3}}c|c|c|c|}
    \hline
    \rowcolor{gray!30} 
    \textbf{Type} & \textbf{Trad. TPI} & \textbf{\lite} & \textbf{Improv. (\%)} \\
    \hline
    \rowcolor{gray!10} 
    \textbf{CC} & 36.28 & 11.26 & \cellcolor{green!10}69\% \\ \hline
    \rowcolor{gray!5} 
    \textbf{Obs} & 32.53 & 20.02 & \cellcolor{green!10}38\% \\ \hline
    \rowcolor{gray!10} 
    \textbf{CC+Obs} & 36.28 & 31.28 & \cellcolor{green!10}14\% \\
    \hline
    \end{tabular}
    \scriptsize
    \raggedright\\ 
    \vspace{0.05cm}{\textbf{CC:} Controllability $\Leftrightarrow$ \lite~: \{Config1/Config2\}\_CC \\
    \textbf{Obs:} Observability $\Leftrightarrow$ \lite~: \{Config1/Config2\}\_Obs \\
    \textbf{CC + Obs:} Combined $\Leftrightarrow$ \lite~: \{Config1/Config2\}\_Obs\_CC}
\end{minipage}
}
\caption{Gate/FF area for SkyWater 130nm std. cell lib. (left) and area comparison between traditional TPI and the proposed \lite~approach (right).}
\label{tab:area_comparison_tpi}
\vspace{-2em}
\end{table}

\begin{figure*}[!htbp]
\centering
\subfloat[Fault Coverage for s13207 using Random Test Patterns.]{\includegraphics[width=0.9\columnwidth]{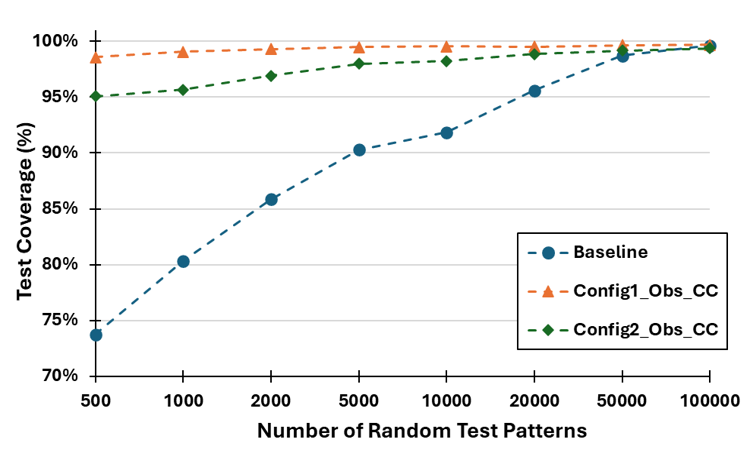}
\label{fig:RP_s13207}}
\hfill
\subfloat[Fault Coverage for s38417 using Random Test Patterns.]{\includegraphics[width=0.9\columnwidth]{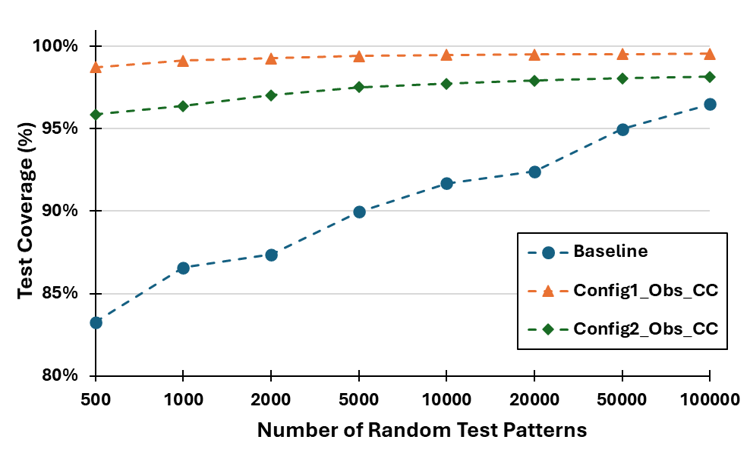}
\label{fig:RP_s38417}}
\hfill
\subfloat[Fault Coverage for b19 using Random Test Patterns.]{\includegraphics[width=0.9\columnwidth]{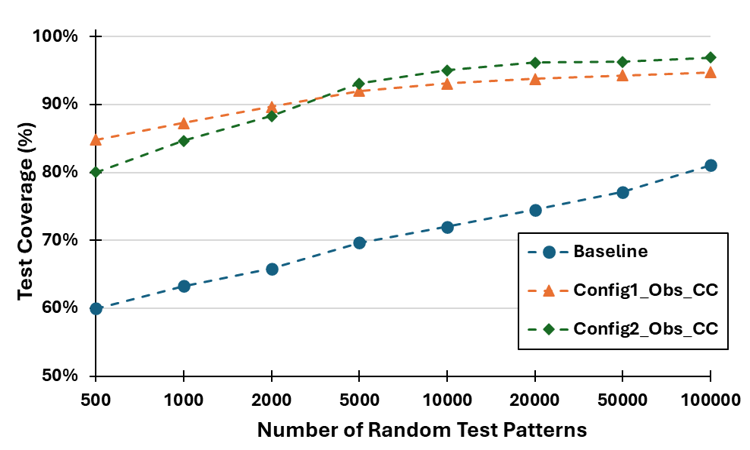}
\label{fig:RP_b19}}
\hfill
\subfloat[Fault Coverage for b22 using Random Test Patterns.]{\includegraphics[width=0.9\columnwidth]{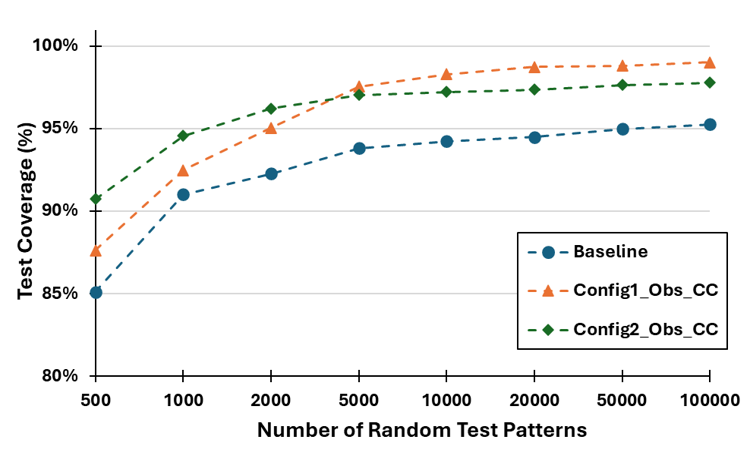}
\label{fig:RP_b22}}
\caption{Fault coverage using random test patterns for ISCAS89 and ITC99 benchmarks: (a) s13207, (a) s38417, (c) b19, and (d) b22. The fault coverage with randomly generated test patterns improves significantly in \lite~due to its ability to sensitize random-resistant faults, compared to the baseline ATPG.}%
\label{fig:random_patterns}%
\end{figure*}

\subsection{Area Comparison with State-of-the-art TPI Techniques}

Test points refer to the points/nets where the commercial EDA tools (e.g., Synopsys TestMAX) insert additional logic to improve the testability of the design. The primary goal of TPI is to increase the observability and controllability of internal nodes within a circuit, which are otherwise difficult to access directly during testing. Fig. \ref{fig:TP_cc0} and \ref{fig:TP_cc1} demonstrate control\_0 or control\_1 test points, respectively, which allow hard-to-control signals to be controllable for some test vectors for testability improvement.  Fig. \ref{fig:TP_co} illustrates observable test point insertion, which is typically inserted at hard-to-observe signals in a design to reduce the number of test patterns and increase coverage.

State-of-the-art TPI techniques \cite{tpi,tpi_touba} lead to an increase in design area due to additional logic and can potentially impact the critical path due to the additional functional FFs, resulting in extra routing. In our proposed approach, \lite~incorporates XOR and MUX for observability improvement and only MUX for controllability improvement without adding any additional sequential elements. The combinational logic introduced by \lite~adds in additional area and delay overhead, but it doesn't impact the critical path delay. Table \ref{tab:area_comparison_tpi} provides a comparative analysis of additional area overhead incurred by state-of-the-art TPI and proposed \lite~approach for improving controllability, observability, and both calculated using cell area values for different gates and FFs from Skywater 130nm\footref{skywater} std. cell library. We observe that \lite~substantially outperforms TPI techniques in terms of area overhead, making it a more practical and efficient solution for testability improvement for larger designs.

\vspace{-1em}

\subsection{Improved Test Coverage with Random Patterns}

In our analysis of fault coverage using varying numbers of random test patterns generated by ATPG, we compared the baseline design utilizing traditional scan flip-flops with \lite-inserted designs. Figure \ref{fig:random_patterns} illustrates the test coverages achieved for varying numbers of test patterns, ranging from 500 to 100K, across four different-sized benchmarks with two from ISCAS89 (Fig. \ref{fig:RP_s13207} and \ref{fig:RP_s38417}) and two from ITC99 (Fig. \ref{fig:RP_b19} and \ref{fig:RP_b22}). 
We observed a significant improvement in fault coverage for \lite-inserted designs with both \textbf{Config1\_Obs\_CC} and \textbf{Config2\_Obs\_CC} configurations outperforming the baseline. Notably, more than 95\% coverage was achieved with approximately 50,000 test patterns, while the baseline design could only achieve a lower coverage within the same number of patterns. As the number of patterns increased, the coverage approached 100\%, which highlights the efficacy of \lite~in improving testability. This improvement in fault coverage in \lite~also indicates the minimization of the impact by RPR faults, which significantly impacts the coverage in traditional scan-based designs or modified designs using TPI techniques. Moreover, the essential observation is the higher coverage achieved with a lower number of test patterns, which implies a reduction in the test cycles required for complex designs. This reduction is particularly beneficial for large-scale designs, as it significantly improves testing efficiency without compromising fault coverage. 

\vspace{-1em}

\subsection{Performance Analysis for Varying XOR Sizes}

\begin{figure}[!htbp]
\centering
\subfloat[Improving \#Patterns ($\upuparrows$) and Increasing Area Overhead ($\downdownarrows$) vs XOR Size ($2-5$) for s13207 benchmark.]{\includegraphics[width=\columnwidth]{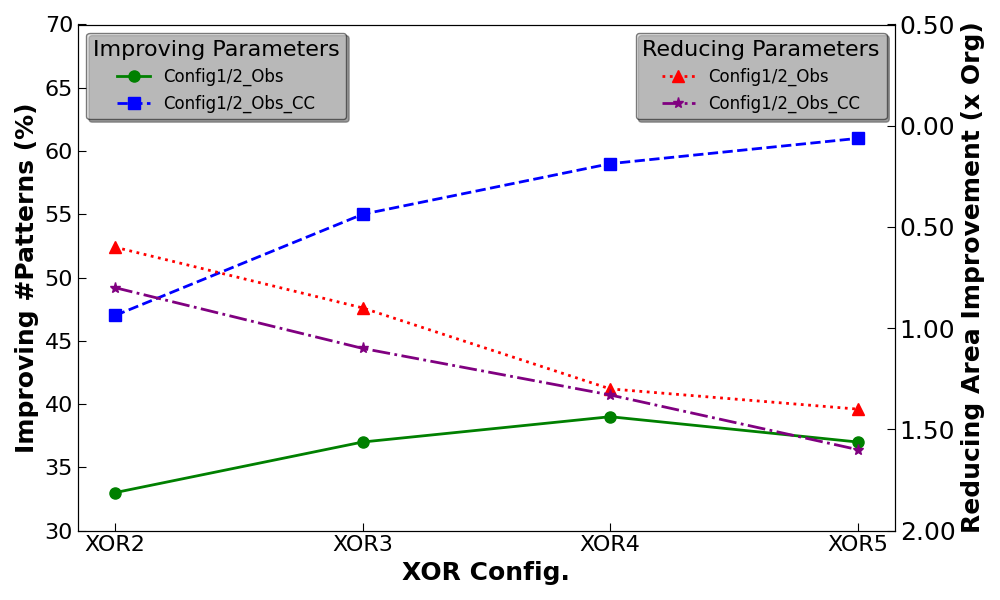}
\label{fig:xor_s13207}}
\hfill
\subfloat[Improving \#Patterns ($\upuparrows$) and Increasing Area Overhead ($\downdownarrows$) vs XOR Size ($2-5$) for s38417 benchmark.]{\includegraphics[width=\columnwidth]{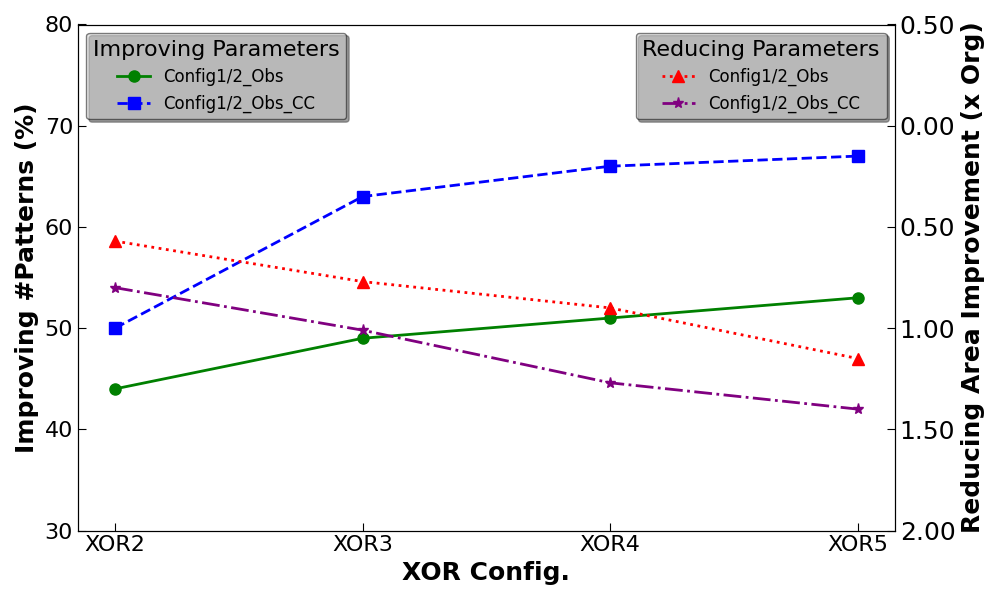}
\label{fig:xor_s38417}}
\caption{Performance analysis for varying sizes of XOR up to 5 inputs for ISCAS89 benchmarks: (a) s13207, and (b) s38417.}%
\label{fig:xor_comparison}
\vspace{-1em}
\end{figure}

We analyzed the performance of XOR gates by varying the input size from 2 to 5, aiming to comprehend how increasing input sizes impact testability analysis. As the input size grows, we can include more low observable nets in the XOR inputs of the proposed LITE configurations, allowing for increased net consideration to improve testability. Fig. \ref{fig:xor_comparison} shows that increasing the XOR input size contributes to reducing the number of test patterns, implying the reduction in the number of test cycles needed without impacting the fault coverage for two ISCAS89 benchmarks (s13207 and s38417). However, this reduction in test patterns comes at the cost of increased area overhead, as larger input sizes lead to a higher area footprint. Notably, we observed that beyond XOR$5$, further increases in input size result in saturation in terms of pattern improvement percentage. This highlights a trade-off between achieving higher pattern improvement and managing area overhead. Consequently, designers must carefully balance these factors when selecting the optimal XOR gate size based on the specific requirements and available resources. Our proposed automated toolflow offers flexibility to allow designers to specify the desired XOR size for the respective \lite~configuration and generate LITE-inserted netlists for efficient testability analysis.

For our analysis, we selected XOR$2$ since it offers considerable pattern improvement ($\approx$ 30\%) while incurring the lowest area overhead ($\approx$ 1.5 $\times$ org) compared to larger sizes of XOR. 

\section{Discussion}
\label{sec:discuss}
In this section, we discuss avenues to further improve the hardware overhead in \lite~and future research directions with respect to its extension in other areas. 

\vspace{-1em}
\subsection{Further Improvement of \lite~Overhead}
While \lite~improves the overhead compared to traditional test point insertion approaches, there are several opportunities to further reduce the hardware overhead. These include the following two solutions. First, the multiplexer and XOR logic inserted for a scan flip-flop for observability improvement can be integrated into the scan chain and custom-designed as a library cell. We anticipate 30-40\% area reduction of this logic through custom design based on our analysis of active area (W$\times$L of the transistors) and layout area of the SkyWater 130nm library. Second, the current implementation of \lite~aims at augmenting most of the scan flip-flops (except for the ones for which specific criteria, such as inadequate number of low-observable or low-controllable nodes) with the additional testability-enhancement logic. While this approach is effective in improving testability, an optimal choice of SFFs for the proposed instrumentation can entail further reduction in overhead while maintaining the testability improvement. This can be accomplished through an iterative approach, where the SFF selection and test generation run in a loop, and the selection of a specific SFF is guided by the impact on testability. It is also possible to formulate the SFF selection problem as a combinatorial optimization (CO) problem and solve it with an appropriate CO solver. 

\vspace{-1em}
\subsection{Extension of \lite}
The proposed approach can be extended in many ways. First, the additional logic for improving controllability can be used to minimize test power during scan shift operation. This can be achieved by controlling the multiplexer at the out of SFF to disable switching in combinational logic. Second, \lite~can be attractive for improving the test efficiency of chiplet-based 3D ICs, where high-density I/O creates significant test challenges with respect to managing thermal issues and poor structural test coverage. The higher density of I/O and the addition of 3D interconnects usually necessitate a greater number of TPs, which can increase test time and complexity. Since \lite~can greatly enhance both ATPG and random pattern testability, as shown in Fig. \ref{fig:random_patterns}, it can be used in the chiplets as a low-overhead alternative to TPs. Further, it can assist 3D IC BIST infrastructure in achieving efficient self-testing of the chiplets as well as inter-chiplet communications. It can also be integrated with scan protection techniques like \cite{bhunia2024invisible} for improved scan chain security with test efficiency. We plan to explore possible extensions of \lite~in the directions noted above in our future work.   

\vspace{-1em}
\section{Conclusion}
\label{sec:conclude}

We have presented \lite, a lightweight scan instrumentation technique that incorporates additional logic into functional scan flip-flops to achieve significant improvement in the testability of a design. It connects a set of select internal nodes to the scan infrastructure through additional logic and then employs conventional ATPG to generate test patterns more efficiently. We have provided detailed circuit-level implementation for the \lite~approach for both controllability and observability improvement and described algorithmic steps to automatically incorporate the design modifications. We have also shown how \lite~interfaces with commercial ATPG tools (e.g., Synopsys TestMax) and how the proposed design modifications can be implemented using commercial ASIC design tool flow and standard cell library. \lite~does not alter the functionality of the normal mode, nor does it alter the test application process of a design. Using a suite of ISCAS89 and ITC99 benchmarks, we have demonstrated \lite~is effective in significantly improving the number of patterns for ATPG-based testing and coverage for random pattern testing. Compared to TP-based approaches for testability improvement, it reduces the overhead by 14\%--69\%. Our future work will include custom scan cell design for \lite~and optimal SFF selection for scan augmentation to further reduce the hardware overhead, extension of the proposed approach to 3D chiplet-based systems to address its test challenges, and analysis of the \lite~approach efficacy in test power improvement.

\bibliographystyle{IEEEtran}
\bibliography{IEEEabrv,references}

\begin{thebibliography}{10}
\providecommand{\url}[1]{#1}
\csname url@samestyle\endcsname
\providecommand{\newblock}{\relax}
\providecommand{\bibinfo}[2]{#2}
\providecommand{\BIBentrySTDinterwordspacing}{\spaceskip=0pt\relax}
\providecommand{\BIBentryALTinterwordstretchfactor}{4}
\providecommand{\BIBentryALTinterwordspacing}{\spaceskip=\fontdimen2\font plus
\BIBentryALTinterwordstretchfactor\fontdimen3\font minus \fontdimen4\font\relax}
\providecommand{\BIBforeignlanguage}[2]{{%
\expandafter\ifx\csname l@#1\endcsname\relax
\typeout{** WARNING: IEEEtran.bst: No hyphenation pattern has been}%
\typeout{** loaded for the language `#1'. Using the pattern for}%
\typeout{** the default language instead.}%
\else
\language=\csname l@#1\endcsname
\fi
#2}}
\providecommand{\BIBdecl}{\relax}
\BIBdecl

\bibitem{bist}
\BIBentryALTinterwordspacing
E.~McCluskey, ``{Built-In Self-Test Techniques},'' \emph{IEEE Des. Test}, vol.~2, no.~2, p. 21–28, Mar. 1985. [Online]. Available: \url{https://doi.org/10.1109/MDT.1985.294856}
\BIBentrySTDinterwordspacing

\bibitem{lbist}
M.~He, G.~K. Contreras, D.~Tran, L.~Winemberg, and M.~Tehranipoor, ``{Test-Point Insertion Efficiency Analysis for LBIST in High-Assurance Applications},'' \emph{IEEE Transactions on Very Large Scale Integration (VLSI) Systems}, vol.~25, no.~9, pp. 2602--2615, 2017.

\bibitem{tpi}
M.~Williams and J.~Angell, ``{Enhancing Testability of Large-Scale Integrated Circuits via Test Points and Additional Logic},'' \emph{IEEE Transactions on Computers}, vol. C-22, no.~1, pp. 46--60, 1973.

\bibitem{tpi_1}
\BIBentryALTinterwordspacing
B.~Krishnamurthy, ``{A dynamic programming approach to the test point insertion problem},'' in \emph{Proceedings of the 24th ACM/IEEE Design Automation Conference}, ser. DAC '87.\hskip 1em plus 0.5em minus 0.4em\relax New York, NY, USA: Association for Computing Machinery, 1987, p. 695–705. [Online]. Available: \url{https://doi.org/10.1145/37888.38000}
\BIBentrySTDinterwordspacing

\bibitem{tpi_2}
N.~Tamarapalli and J.~Rajski, ``{Constructive multi-phase test point insertion for scan-based BIST},'' in \emph{Proceedings International Test Conference 1996. Test and Design Validity}, 1996, pp. 649--658.

\bibitem{tpi_3}
D.~Das and N.~Touba, ``{Reducing test data volume using external/LBIST hybrid test patterns},'' in \emph{Proceedings International Test Conference 2000 (IEEE Cat. No.00CH37159)}, 2000, pp. 115--122.

\bibitem{tpi_4}
S.~Roy, B.~Stiene, S.~K. Millican, and V.~D. Agrawal, ``{Improved pseudo-random fault coverage through inversions: a study on test point architectures},'' \emph{Journal of Electronic Testing}, vol.~36, pp. 123--133, 2020.

\bibitem{tpi_touba}
J.-S. Yang, B.~Nadeau-Dostie, and N.~A. Touba, ``{Test point insertion using functional flip-flops to drive control points},'' in \emph{2009 International Test Conference}, 2009, pp. 1--10.

\bibitem{ml_tpi_1}
Y.~Sun and S.~Millican, ``{Test Point Insertion Using Artificial Neural Networks},'' in \emph{2019 IEEE Computer Society Annual Symposium on VLSI (ISVLSI)}, 2019, pp. 253--258.

\bibitem{ml_tpi_2}
S.~Millican, Y.~Sun, S.~Roy, and V.~Agrawal, ``{Applying Neural Networks to Delay Fault Testing: Test Point Insertion and Random Circuit Training},'' in \emph{2019 IEEE 28th Asian Test Symposium (ATS)}, 2019, pp. 13--135.

\bibitem{ml_tpi_3}
\BIBentryALTinterwordspacing
Y.~Ma, H.~Ren, B.~Khailany, H.~Sikka, L.~Luo, K.~Natarajan, and B.~Yu, ``{High Performance Graph Convolutional Networks with Applications in Testability Analysis},'' in \emph{Proceedings of the 56th Annual Design Automation Conference 2019}, ser. DAC '19.\hskip 1em plus 0.5em minus 0.4em\relax New York, NY, USA: Association for Computing Machinery, 2019. [Online]. Available: \url{https://doi.org/10.1145/3316781.3317838}
\BIBentrySTDinterwordspacing

\bibitem{ml_tpi_4}
Z.~Shi, M.~Li, S.~Khan, L.~Wang, N.~Wang, Y.~Huang, and Q.~Xu, ``{Deeptpi: Test point insertion with deep reinforcement learning},'' in \emph{2022 IEEE International Test Conference (ITC)}.\hskip 1em plus 0.5em minus 0.4em\relax IEEE, 2022, pp. 194--203.

\bibitem{test_1}
X.~Gu, K.~Kuchcinski, and Z.~Peng, \emph{{Testability analysis and improvement from VHDL behavioral specifications}}.\hskip 1em plus 0.5em minus 0.4em\relax Universitetet i Link{\"o}ping/Tekniska H{\"o}gskolan i Link{\"o}ping. Institutionen f{\"o}r~…, 1994.

\bibitem{test_2}
V.~Chickermane, J.~Lee, and J.~H. Patel, ``{Addressing design for testability at the architectural level},'' \emph{IEEE transactions on computer-aided design of integrated circuits and systems}, vol.~13, no.~7, pp. 920--934, 1994.

\bibitem{test_3}
C.-H. Chen, T.~Karnik, and D.~G. Saab, ``{Structural and behavioral synthesis for testability techniques},'' \emph{IEEE transactions on computer-aided design of integrated circuits and systems}, vol.~13, no.~6, pp. 777--785, 1994.

\bibitem{test_4}
L.~Avra, ``{Allocation and assignment in high-level synthesis for self-testable data paths},'' in \emph{1991, Proceedings. International Test Conference}.\hskip 1em plus 0.5em minus 0.4em\relax IEEE, 1991, p. 463.

\bibitem{test_5}
M.~Nourani and C.~Papachristou, ``{Structural BIST insertion using behavioral test analysis},'' in \emph{Proceedings European Design and Test Conference. ED \& TC 97}.\hskip 1em plus 0.5em minus 0.4em\relax IEEE, 1997, pp. 64--68.

\bibitem{test_6}
S.~Bhattacharya and S.~Dey, ``{H-SCAN: A high level alternative to full-scan testing with reduced area and test application overheads},'' in \emph{Proceedings of 14th VLSI Test Symposium}, 1996, pp. 74--80.

\bibitem{test_7}
W.~Mao and R.~K. Gulati, ``{Improving gate level fault coverage by RTL fault grading},'' in \emph{Proceedings International Test Conference 1996. Test and Design Validity}.\hskip 1em plus 0.5em minus 0.4em\relax IEEE, 1996, pp. 150--159.

\bibitem{vhdl_test}
S.~Boubezari, E.~Cerny, B.~Kaminska, and B.~Nadeau-Dostie, ``Testability analysis and test-point insertion in rtl vhdl specifications for scan-based bist,'' \emph{IEEE Transactions on Computer-Aided Design of Integrated Circuits and Systems}, vol.~18, no.~9, pp. 1327--1340, 1999.

\bibitem{podem}
P.~Goel and B.~C. Rosales, ``{PODEM-X: An automatic test generation system for VLSI logic structures},'' in \emph{Proceedings of the 18th Design Automation Conference}, ser. DAC '81.\hskip 1em plus 0.5em minus 0.4em\relax IEEE Press, 1981, p. 260–268.

\bibitem{test_gen_1}
\BIBentryALTinterwordspacing
F.~E. Norrod, ``{An automatic test generation algorithm for hardware description languages},'' in \emph{Proceedings of the 26th ACM/IEEE Design Automation Conference}, ser. DAC '89.\hskip 1em plus 0.5em minus 0.4em\relax New York, NY, USA: Association for Computing Machinery, 1989, p. 429–434. [Online]. Available: \url{https://doi.org/10.1145/74382.74454}
\BIBentrySTDinterwordspacing

\bibitem{scan_reorder_1}
Y.-Z. Wu and M.~C.-T. Chao, ``Scan-chain reordering for minimizing scan-shift power based on non-specified test cubes,'' in \emph{26th IEEE VLSI Test Symposium (vts 2008)}, 2008, pp. 147--154.

\bibitem{scan_reorder_2}
D.~Ghosh, S.~Bhunia, and K.~Roy, ``A low-complexity scan reordering algorithm for low power test-per-scan bist,'' in \emph{Proceedings of the International Conference on VLSI Design, Mumbai, India}.\hskip 1em plus 0.5em minus 0.4em\relax Citeseer, 2004, pp. 5--9.

\bibitem{scan_reorder_3}
S.~Bhunia, H.~Mahmoodi, D.~Ghosh, and K.~Roy, ``Power reduction in test-per-scan bist with supply gating and efficient scan partitioning,'' in \emph{Sixth international symposium on quality electronic design (isqed'05)}, 2005, pp. 453--458.

\bibitem{scan_locking_1}
D.~Zhang, M.~He, X.~Wang, and M.~Tehranipoor, ``{Dynamically obfuscated scan for protecting IPs against scan-based attacks throughout supply chain},'' in \emph{2017 IEEE 35th VLSI Test Symposium (VTS)}, 2017, pp. 1--6.

\bibitem{scan_locking_2}
S.~Paul, R.~S. Chakraborty, and S.~Bhunia, ``{VIm-Scan: A Low Overhead Scan Design Approach for Protection of Secret Key in Scan-Based Secure Chips},'' in \emph{25th IEEE VLSI Test Symposium (VTS'07)}, 2007, pp. 455--460.

\bibitem{bhunia2024invisible}
S.~Bhunia, P.~D. Gaikwad, J.~W. Cruz, and S.~Paria, ``Invisible scan architecture for secure testing of digital designs,'' Apr.~9 2024, uS Patent 11,953,548.

\bibitem{scoap}
L.~Goldstein and E.~Thigpen, ``{SCOAP: Sandia Controllability/Observability Analysis Program},'' in \emph{17th Design Automation Conference}, 1980, pp. 190--196.

\end{thebibliography}

\end{document}